\documentclass[twocolumn,superscriptaddress,pre]{revtex4}

\usepackage{amsmath}
\usepackage{amsbsy}
\usepackage{amssymb}
\usepackage{mathtools}
\usepackage{dcolumn}
\usepackage{enumitem}
\usepackage{graphicx}
\usepackage{float}
\usepackage{xcolor}
\usepackage{bigints}
\usepackage{cleveref}
\usepackage{listings}
\usepackage{tcolorbox}
\definecolor{codegreen}{rgb}{0,0.6,0}
\definecolor{codegray}{rgb}{0.5,0.5,0.5}
\definecolor{codepurple}{rgb}{0.58,0,0.82}
\definecolor{backcolour}{rgb}{0.95,0.95,0.92}

\lstdefinestyle{mystyle}{
    backgroundcolor=\color{backcolour},
    commentstyle=\color{codegreen},
    keywordstyle=\color{magenta},
    numberstyle=\tiny\color{codegray},
    stringstyle=\color{codepurple},
    basicstyle=\footnotesize\ttfamily,
    breakatwhitespace=false,
    breaklines=true,
    captionpos=b,
    keepspaces=true,
    numbers=left,
    numbersep=5pt,
    showspaces=false,
    showstringspaces=false,
    showtabs=false,
    tabsize=2,
    mathescape=true, 
}

\tcbuselibrary{listingsutf8}
\newtcblisting{algorithm}[1][]{
    listing only,
    colback=backcolour,
    colframe=backcolour,
    listing style=mystyle,
    boxsep=0pt, 
    top=4pt, 
    bottom=4pt, 
    #1
}
    
\crefname{algocfline}{alg.}{algs.}
\Crefname{algocfline}{Algorithm}{Algorithms}

\newcommand\bm{\boldsymbol}

\begin{document}

\title{An approximation for return time distributions\\of random walks on sparse networks}
\author{Erik Hormann}
 \affiliation{Mathematical Institute, University of Oxford, Oxford, OX2 6GG, United Kingdom}
\author{Renaud Lambiotte}
 \affiliation{Mathematical Institute, University of Oxford, Oxford, OX2 6GG, United Kingdom}
\author{George T. Cantwell}
 \affiliation{Department of Engineering, University of Cambridge, Cambridge, CB2 1PZ, United Kingdom}

\begin{abstract}
    We propose an approximation for the first return time distribution of random walks on undirected networks. We combine a message-passing solution with a mean-field approximation, to account for the short- and long-term behaviours respectively. We test this approximation on several classes of large graphs and find excellent agreement between our approximations and the true distributions.
    While the statistical properties of a random walk will depend on the structure of the network, the observed agreement between our approximations and numerical calculations implies that while local structure is clearly very influential, global structure is only important in a relatively superficial way, namely through the total number of edges.
\end{abstract}

\maketitle

\section{Introduction}

Random walks are fundamental in the field of stochastic processes \cite{feller1991introduction}.  More than a mathematical curiosity, they have been studied and applied in disparate areas, for example, they have been used to model the movements of molecules \cite{wolynes1978molecular}, genetic drift \cite{weinstein2017genetic}, stock prices \cite{godfrey1964random}, and even human decision making \cite{ratcliff2008diffusion}.
A random walk on a discrete space, i.e. a Markov chain, can be conceptualized as occurring on a network.
And conversely, given a network, we can imagine a random walk taking place upon it.
Random walks have the advantage of exploring nonlocal patterns of connectivity allowing us, for example, to capture the influence between pairs of nodes that are not directly connected by an edge. 
This intuition is at the core of several innovative algorithms for analysing networked data \cite{masuda2017random}, such as ranking nodes \cite{brin1998anatomy}, network embedding \cite{perozzi2014deepwalk} or community detection \cite{rosvall2008maps,lambiotte2014random}. Random walks are also an object of study as a simple, linear model for the diffusion of entities on a network, for instance of individuals in metapopulation models in epidemiology \cite{colizza2007reaction}, and are the dual process of popular models for decentralized computation and consensus dynamics \cite{tsitsiklis1986distributed}.

The statistical properties of random walk processes are known theoretically on a range of abstract structures, from lattices to regular trees \cite{redner2001guide}, but their analysis often requires numerical simulations on less regular and more realistic structures, such as complex networks \cite{masuda2017random}.
However, numerical simulations of the random process do not directly uncover or explain its statistical properties.
In this paper, we focus on the first return time distributions, capturing the statistical properties of the time it takes for a walker to return to a node.
This is a fundamental property of random walks that has been studied extensively in the field, and provide useful information on the dynamical random processes that generate them \cite{masuda2004return, kujawski2007preferential}.
In fact, full knowledge of the return time distribution is stronger than knowledge of the eigen-spectrum, and thus may be sufficient to fully determine a network \cite{haemers2016almost}.

The main contributions of this work are a set of formulas to approximate first return time distributions on networks.
Our formulas can be solved computationally efficiently and depend on the local structure around nodes, as well as the total number of edges -- a simple measure of global structure.
We verify on different families of graphs that the approximate expressions accurately describe random walks on networks, even in situations with a high density of short cycles.

The remainder of this paper is structured as follows.
First, we give a gentle introduction to random walks, return times, and generating functions.
Next, we review existing approximations: heterogeneous mean field approximations and tree approximations, for the quantities of interest.
We then introduce a combined approximation and demonstrate its accuracy on large locally tree-like networks.
Finally, we improve the approximation---built on the assumption of a locally tree-like network---to account for cycles.
Our final approximation appears to be accurate on sparse networks with arbitrary local topology.
We test our approximations by comparing against direct numerical calculations and we conclude with a discussion of the implications of our results.

\section{Basics of random walks and networks}\label{sec:intro}
We consider a random walk process on an undirected, unweighted network $G$ composed of $n$ nodes and $m$ edges. 
When a random walker is at node~$i$, it jumps to one of the neighbours of  $i$ chosen at random.
Using the adjacency matrix
\begin{equation}
	A_{ij} =
	\begin{cases}
		1 \quad \text{if edge }(i,j) \in G \\
		0 \quad \text{otherwise}
	\end{cases}
\end{equation}
and denoting the degree (total number of neighbours) of node $i$ as $k_i = \sum_j A_{ij}$, we can write the probability of moving from $i$ to $j$ as
\begin{equation}
	W_{ij} = \frac{A_{ij}}{k_i}.
\end{equation}
After $t$ time steps, the probabilities of moving from one node to another are encoded in the entries of the matrix  $W^t$.
If the initial location of the random walker is given by the distribution $\bm{p}$,  the distribution of walkers after $t$ time steps is thus given by $\bm{p} W^t$.

Assuming that the network is connected (which we will  assume henceforth), there is a unique stationary distribution $\bm{\pi} = \bm{\pi} W$.
The stationary distribution is given by $\pi_i = k_i / 2m$, with a proper normalisation ensured by $\sum_i k_i=2m$, where $m$ is the total number of edges in the network. Stationarity can be verified easily with:
\begin{equation}
	\sum_{i} \pi_i W_{i j} = \sum_i \frac{k_i}{2m} \frac{A_{ij}}{k_i} = \frac{\sum_i A_{ij} }{2m} = \frac{k_j }{ 2m} = \pi_j.
\end{equation}
The proof for uniqueness is similarly simple but longer, and we refer to \cite{norris1998markov}. 
After a sufficiently large number of time steps, the distribution of the walker will converge to $\bm{\pi}$, regardless of where it started (unless the network is bipartite, but we will not consider this special case here).

One quantity of interest is the return probability, $X_i(t)$, defined as the probability that a random walk that starts at node~$i$ will be back at $i$ after $t$ steps.
In terms of the matrix $W$ we have
\begin{equation}
	X_i(t) = (W^t)_{ii}.
\end{equation}
A closely related quantity of interest is the \emph{first return time probability} $Y_i(t)$, also known as \emph{recurrence time probability}, now defined as the probability that a random walk that starts at node~$i$ returns to $i$ \emph{for the first time} at time $t$.
On a finite network, a random walk must return at some point, and so we have $\sum_{t=1}^{\infty} Y_i(t) = 1$ for any node $i$.

The distribution of first return times, $Y_i(t)$, generally depends on the whole structure of the network.
However, its average value has a simple form, known as the Kac formula, which can be derived as follows \cite{levin2017markov}.
Let $\mu({j \leadsto i})$ be the average time for a random walk that starts at $j$ to reach $i$, in which case, $\mu({i \leadsto i}) = \sum_{t} t\, Y_i(t)$ is the average first return time.
A walker that starts at $j$ will make its first step to $k$ with probability $W_{jk}$.
If $k=i$ then we are done, and we have moved from $j \leadsto i$ in a single step.
Otherwise, we will have  moved one step but will now be at some other node $k\neq i$.
The expected additional time to reach $i$ is then $\mu(k\leadsto i)$.
Writing this out gives
\begin{align}
	\mu({j \leadsto i}) &= 1 + \sum_{k \neq i} W_{jk} \, \mu(k \leadsto i) \nonumber \\
	 &= 1 + \sum_{k} W_{jk} \, \mu(k \leadsto i) - W_{ji} \, \mu(i \leadsto i).
\end{align}
If we multiply both sides of this equation by $k_j$, sum over $j$, and use the fact that 
$\sum_j k_j W_{ji} = k_i$ we find Kac formula
\begin{equation}
	\mu( i \leadsto i) = \sum_{t=1}^{\infty} t\, Y_i(t) =  \frac{2m}{k_i},
\end{equation}
and the a priori striking result that the average first return time of a node only depends on its degree.

As mentioned above, the quantities $X_i(t)$ and $Y_i(t)$ are closely related.
Recall that $X_i(t)$ is the probability that a walk that started at $i$ is back at $i$ at after $t$ steps.
In contrast, $Y_i(t)$ is the probability that a walk that started at $i$ is back at $i$ \emph{for the first time} after $t$ steps.
To find the relation between these  quantities, we consider the probability that a walker  starting at $i$ has returned to $i$ for the $q$th time after $t$ steps.
In this scenario, there are $q$ walks that started and ended at $i$ for the first time.
Each of these would have taken $t_s$ steps with probability $Y_i(t_s)$, subject to the constraint $\sum_{s=1}^{q} t_s = t$ so that the total number of steps for all $q$ walks is equal to $t$.
The probability for a walker to return for the $q$th time after $t$ steps is thus
\begin{equation}
	\sum_{t_1=1}^{\infty} \dots \sum_{t_q=1}^{\infty} Y_i(t_1)\, \dots\, Y_i(t_q)\, \delta\big(t, {\textstyle{\sum_{s=1}^{q}}} t_s\big),
\end{equation}
where $\delta$ is the Kronecker delta.
The quantity $X_i(t)$ is the probability that a walk has returned to $i$ any number of times, and so summing the above quantity over $q$ we can write 
\begin{equation}
	X_i(t) = \sum_{q=1}^{\infty} \left(\, \sum_{t_1=1}^{\infty} Y_i(t_1) \dots \sum_{t_q=1}^{\infty} Y_i(t_q)\, \delta\big(t, {\textstyle{\sum_{s}}} t_s\big)\right)
	\label{eq:X_to_Y}
\end{equation}
for $t\geq1$.

At first glance, Eq.~\eqref{eq:X_to_Y} does not appear to be a particularly useful relation, but it becomes clearer when considering \emph{generating functions} of the process.
Let
\begin{equation}
	R_i(z) = \sum_{t=0}^{\infty} z^t\, X_i(t)
	\label{eq:R_defn}
\end{equation}
and
\begin{equation}
	F_i(z) = \sum_{t=0}^{\infty} z^t\, Y_i(t).
\end{equation}
be the generating functions for the return time and the first return times.
By taking derivatives of the generating functions, we can recover the underlying distributions, or compute different quantities of interest.
For example, the derivative
\begin{equation}
	F_i'(1) = \sum_{t=1}^{\infty} t\, Y_i(t) = \mu( i \leadsto i)
\end{equation}
is the mean first return time.
Likewise, the variance of the first return time can be computed from the second derivative.

Inserting Eq.~\eqref{eq:X_to_Y} into the definition of $R_i(z)$ we get
\begin{align}
	R_i(z) &= 1 + \sum_{q=1}^{\infty} \left( \sum_{t=1}^{\infty} z^t \, Y_i(t) \right)^q \nonumber \\
	&=  1 + \sum_{q=1}^{\infty} F_i(z)^q \nonumber \\
	&= \frac{1}{1-F_i(z)}. \label{eq:FRTD2MGF}
\end{align}
This relation shows that if we know the generating function for either $X_i(t)$ or $Y_i(t)$, then we know the generating function for the other.
In general, the exact form for these distributions has an intricate dependence on the structure of the network. It is the purpose of the next section to derive  approximations that may help us to identify the important structural features determining first return times properties, and also to estimate them in large-scale networks.

\section{Approximations}
\subsection{Heterogeneous mean-field approximation}\label{sec:mf_approx}
As a first approximation, we consider a heterogeneous mean-field approximation where the walker may jump, at each step, to any node $i$ (including where it came from) with a probability given by the stationary distribution for the random walk $\pi_i = k_i/2m$. 
In this approximation, the process relaxes to  the stationary distribution in one single step, and it can be seen as a random walk on a weighted network (with self-loops) with adjacency matrix $k_i k_j/2m$. The return probability is then given by
\begin{equation}
	\widetilde{X}_i(t) = \begin{cases}
		1 \quad &\text{for } t=0 \\
		{k_i / 2m} \quad &\text{for } t \geq 1,
	\end{cases}
\end{equation}
where we placed a tilde over the $X$ (i.e., $\widetilde{X}$) to mark the approximation, and the assumption is that
\begin{equation}
    X_i(t) \simeq \widetilde{X}_i(t).
\end{equation}

The corresponding first return times, $\widetilde{Y}_i(t)$, can be computed as follows.
For the walker to return \emph{for the first time} after $t$ steps, it must have not returned for the previous $t-1$, and then return at $t$.
If the walker returns at each time step with probability $\pi_i = k_i / 2m$ then we have
\begin{equation}
	\widetilde{Y}_i(t) = \left( 1 -  \pi_i \right)^{t-1} { \pi_i }
\end{equation}
and
\begin{equation}
	F_i(z) \simeq \widetilde{F}_i(z) = \sum_{t=0}^{\infty} z^t\, \widetilde{Y}_i(t) = \frac{z \pi_i }{1 - z + z \pi_i}.
\end{equation}
This approximation is a crude one and may provide wrong estimates in practice, but it satisfies two important properties:
\begin{equation}
	\sum_{t=1}^{\infty} \widetilde{Y}_i(t) = \widetilde{F}_i(1) = 1
\end{equation}
and
\begin{equation}
	\sum_{t=1}^{\infty} t\, \widetilde{Y}_i(t) = \widetilde{F}_i'(1) = \frac{2m}{k_i}
\end{equation}
both of which must be true on a finite network.

\subsection{Tree approximation}\label{sec:tree_approx}

Another approximation can be obtained by assuming that the network has no cycles, i.e., it is a tree.
Recall that $Y_j(t)$ is the probability that a random walk that starts at node~$j$ will return to node~$j$ for the first time after $t$ steps.
Let us define the related probability $Y_{i \leftarrow j}(t)$ to be the probability that such a walk occurs with the additional condition that it never visited node~$i$, i.e., the probability that a random walk that starts at $j$ returns to $j$ for the first time after $t$ steps and does not visit $i$.
We can relate the quantities as follows.
If a random walk starts at node~$i$, its first step will be to one of $i$'s neighbours $j$, with probability $1/k_i$.
Now that the walk is at $j$ it could immediately return to $i$ with probability $1/k_j$.
Thus, the probability of returning after exactly two steps is
\begin{equation}
	Y_i(2) = \sum_{j \in N_i} \frac{1}{k_i k_j}.
\end{equation}
If the walk does not return in two steps, it will leave $j$ to visit one of $j$'s neighbours.
From now on, we assume that the graph is a tree, looking for the approximation of $Y_i$ denoted by $\widetilde{Y}_i$.
Under this tree assumption, the walk cannot get back to node~$i$ without revisiting $j$.
This means that there must be some number $q \geq 0$ of excursions from $j$ that never reach $i$.
Denote the probability distribution for the length of these excursions that start and end at $j$, and never visit $i$ as $\widetilde{Y}_{i \leftarrow j}(t)$.
Then, the probability that we return to node~$i$ for the first time after exactly $t$ steps is
\begin{align}
	\widetilde{Y}_i(t) = \sum_{j \in N_i} \frac{1}{k_i k_j} \sum_{q=0}^{\infty} &  ~\sum_{t_1} \dots  \sum_{t_q} 
 \Big( \widetilde{Y}_{i \leftarrow j}(t_1) \widetilde{Y}_{i \leftarrow j}(t_2) \dots \nonumber \\ 
 & \dots \widetilde{Y}_{i \leftarrow j}(t_q)\,  \delta \big( t-2, {\textstyle\sum_s} t_s \big) \Big).
	\label{eq:Y_i_MP}
\end{align}
Multiplying Eq.~\eqref{eq:Y_i_MP} by $z^t$ and summing gives
\begin{align}
	\widetilde{F}_i(z) &= \sum_{t=0}^{\infty} z^t\, \widetilde{Y}_i(t)  \nonumber \\
	&= \sum_{j \in N_i} \frac{z^2}{k_i k_j} \sum_{q=0}^{\infty} \left( \sum_{t=0}^{\infty} z^t \, \widetilde{Y}_{i \leftarrow j}(t) \right)^q  \nonumber \\
	&= \sum_{j \in N_i} \frac{z^2}{k_i k_j} \left( \frac{1}{1 - \widetilde{F}_{i \leftarrow j}(z) }\right),
	\label{eq:F_i_MP}
\end{align}
where $\widetilde{F}_{i \leftarrow j}(z) = \sum_{t=0}^{\infty} z^t \, \widetilde{Y}_{i \leftarrow j}(t)$.
This relates the generating functions for $\widetilde{Y}_i(t)$ to those of $\widetilde{Y}_{i \leftarrow j}(t)$.

To make further progress we can make an analogous argument to compute $\widetilde{Y}_{i \leftarrow j}(t)$.
In this case, the walk must first visit a neighbour of $j$ that is not $i$, and then return to $j$ at a later step.
Following through the same line of argument leads to
\begin{equation}
	\widetilde{F}_{i \leftarrow j}(z) = \sum_{k \in N_j \setminus i} \frac{z^2}{k_j k_k} \left( \frac{1}{1 -  \widetilde{F}_{j \leftarrow k}(z) } \right),
	\label{eq:tree_MP}
\end{equation}
where the sum over $k \in N_j \setminus i$ is a sum over all neighbours of $j$ except for $i$.
The solution to this set of equations can be inserted into Eq.~\eqref{eq:F_i_MP} to find the generating functions to the first return times.

Equation~\eqref{eq:tree_MP} can be evaluated on any given network to make predictions.
However, we may also ask about the distribution of solutions.
On a locally tree-like network with degree distribution $\rho_k$ and mean degree $\langle k \rangle = \sum_k k \rho_k$, there will be a distribution of solutions.
For a given value of $z$, let $P_{\leftarrow}^{(z)}(F, k)$ be the probability density for a randomly chosen message to take value $F$ and lead to a node of degree $k$.
This probability density solves the self-consistent equation
\begin{align}
P_{\leftarrow}^{(z)}(F,k) &= \nonumber \\
\frac{k \rho_k}{\langle k \rangle} &\bigintssss \prod_{j=1}^{k} \mathrm{d}P_{\leftarrow}^{(z)}(F_j, k_j) 
\delta \bigg( {\sum_{j=1}^{k}} \frac{z^2}{k k_j(1- F_j)} - F \bigg)	\label{eq:pop_dyn_message}
\end{align}
which can be solved numerically using the population dynamic algorithm \cite{mezard2009information}, see Appendix~\ref{sec:pop_dyn_alg}.

On the $d$-regular tree, each node has degree $d$.
In this case, Eqs.~\eqref{eq:F_i_MP} and~\eqref{eq:tree_MP} are the same for all $i,j$, leading to
\begin{equation}
	\widetilde{F}(z) =  \frac{z^2}{d} \left( \frac{1}{1 -  \widetilde{F}_{\leftarrow}(z) } \right)
\end{equation}
and
\begin{equation}
	\widetilde{F}_{\leftarrow}(z) =  \frac{z^2 (d-1)}{d^2} \left( \frac{1}{1 -  \widetilde{F}_{\leftarrow}(z) } \right).
\end{equation}
Solving these equations gives
\begin{equation}
	\widetilde{F}(z) = \frac{2 z^2}{d + \sqrt{d^2 + 4 z^2 - 4 d z^2}},
	\label{eq:F_d_reg_tree}
\end{equation}
in agreement with standard results for  random walks on  infinite regular trees \cite{hughes1982random}.

Looking at Eq.~\eqref{eq:F_d_reg_tree} we observe that $\widetilde{F}(1) < 1$ for $d>2$.
This means that the random walk is \emph{transient}: with a nonzero probability, a random walk that starts at $i$ may never return to $i$, or equivalently, $i$ may only be visited a finite number of times.
This property is indeed true on infinite regular trees with $d>2$. However, 
on any finite network, which is our focus,  it is clearly wrong.

\subsection{Combined approximation}

On a locally tree-like graph, the tree approximation is a compelling method to approximate transient behaviour.
However, on a finite network, it makes clear errors, even for basic quantities as  $\widetilde{F}_i(1) \neq 1$ and $\widetilde{F}_i'(1) \neq 2m/k_i$.
Conversely, the mean-field approximation satisfies these conditions but fails to pick up any local structure or transient behaviour.
Can we combine the two, to get a best of both-worlds approximation?

We propose to do this as follows.
First, let $\widetilde{F}_i(z)$ be the solution to the tree approximation, from Eq.~\eqref{eq:F_i_MP}.
Define the quantity
\begin{equation}
	h_i = \frac{2 m  - k_i \widetilde{F}_i'(1) }{ k_i - k_i \widetilde{F}_i(1) }
	\label{eq:tail_mean}
\end{equation}
and then use
\begin{equation}
	F_i(z) = \widetilde{F}_i(z) + \frac{z - z\, \widetilde{F}_i(1)}{z + h_i - z h_i}.
	\label{eq:finite_approx}
\end{equation}
For small values of $t$, the first term in this expression will be dominant, because $h_i$ is large.
In other words, this approximation makes similar predictions to the tree approximation at short times, but additionally satisfies the required properties:
\begin{equation}
	F_i(1) = 1 \quad\quad \text{and} \quad\quad F_i'(1) = 2m/k_i
\end{equation}
as can easily be checked. Eq.~\eqref{eq:finite_approx} thus defines a distribution for a recurrent process with the correct mean and a reasonable approximation at short times on locally tree-like graphs.

\begin{figure}
  \includegraphics[width=0.45\textwidth]{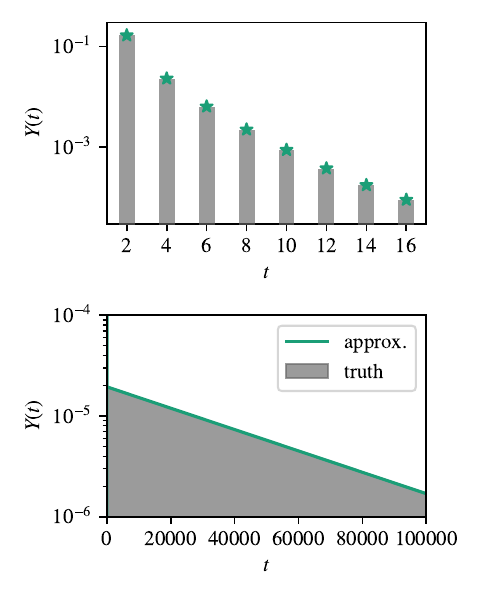}
  \caption{First return time distribution, $Y(t)$, for random walks on a $6$-regular random graph with $n=2^{15}$ nodes. The ground truth is computed for an arbitrary node with matrix iteration. The approximation is computed from derivatives of Eq.~\eqref{eq:finite_random_reg_genfun} at $z=0$.
  Note, the tail of the distribution is straight on a semi-log plot, indicating exponential decay.
  The slope of this tail on a logarithmic scale is the inverse of the quantity we call the tail-mean, approximated by Eq.~\eqref{eq:tail_mean}.
  }
  \label{fig:rand_reg}
\end{figure}

\subsection{Numerical experiments for locally tree-like graphs}\label{sec:tree_sims}

In this section, we test the validity of the combined approximation on a range of networks. 

A first simple example is  the random $d$-regular graph.
In this case we have 
\begin{equation}
	h = \frac{\left(d-1\right) \left(n d - 2 n - 2 \right)}{\left(d-2\right)^2}
\end{equation}
and then
\begin{equation}
	F(z) = \frac{2 z^2}{d + \sqrt{d^2 + 4 z^2 - 4 d z^2}} + \frac{z - z/(d-1)}{z + h - z h}.
	\label{eq:finite_random_reg_genfun}
\end{equation}
In Fig.~\ref{fig:rand_reg} we compare this approximation to random regular graphs with $n=2^{15}$ nodes and degree $d=6$.  We find excellent agreement between the predictions of Eq.~\eqref{eq:finite_random_reg_genfun} and simulations.

\begin{figure}
  \includegraphics[width=0.45\textwidth]{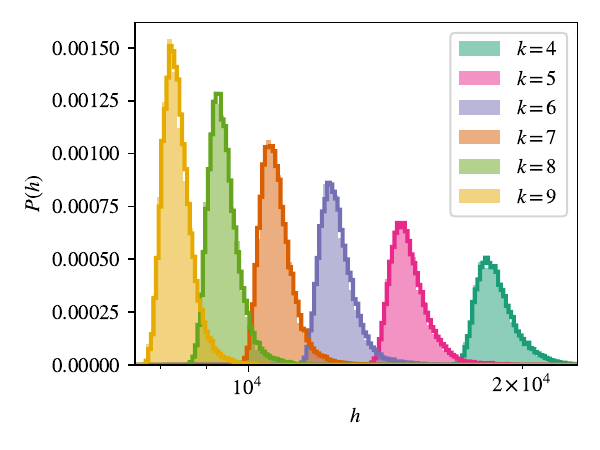}
	\caption{Slopes of the tails of return time distributions. The filled histogram shows the distributions of tails slopes for nodes with degree $k$ in a Poisson random graph with $10^5$ nodes and mean degree $6$.  The lines show the same quantity, but as predicted by Eqs.~\eqref{eq:pop_dyn_message} and~\eqref{eq:tail_mean}.}
  \label{fig:poisson_tails}
\end{figure}

Second, we explore the validity of the equation for the \emph{distribution} of messages, Eq.~\eqref{eq:pop_dyn_message}, on locally tree-like graphs.
To this end, we generated networks from the Poisson random graph with $n=10^5$ and mean degree $6$.
Having generated the networks, we computed the first return time distributions by iterating the random walk matrix.
These first return time distributions have exponential tails, and so for each node in these networks we numerically computed exponential rate of decay -- the slope of the line on a log-linear plot, see e.g. Fig.~\ref{fig:rand_reg}(b).
To compare these results against theoretical predictions, we also solved Eq.~\eqref{eq:pop_dyn_message} for Poisson distributed degree with mean degree $6$. 
This was done using a standard population dynamics algorithm (see Appendix A).
This provides a solution for the \emph{distribution} of messages.
Recall, the message passing solution is wrong at long times -- it predicts a transient random walk. Nevertheless, we can insert randomly chosen messages into the equation for $h$ to predict the slope of the exponential tail of a randomly chosen node.
Figure~\ref{fig:poisson_tails} shows both the empirical and theoretical the distribution of tail slopes, again demonstrating excellent agreement of the theory – not only on average, but in distribution.
Interestingly, the predictions of this approximation do not change regardless of large scale structure, such as community structure, or core-periphery structure \cite{gallagher2021clarified}.

For example, a stochastic block model with $c$ equally sized communities is locally tree-like if $c$ does not grow with $n$ \cite{newman2018networks, decelle2011inference}.
Thus, our approximation predicts that, even when the community structure is extremely strong, random walks will appear to behave on the block model as though they are on a random graph for sufficiently large $n$.
To verify this prediction, we compared the tree-like predictions to ground-truth on stochastic block models of increasing size.

\begin{figure}
  \includegraphics[width=0.45\textwidth]{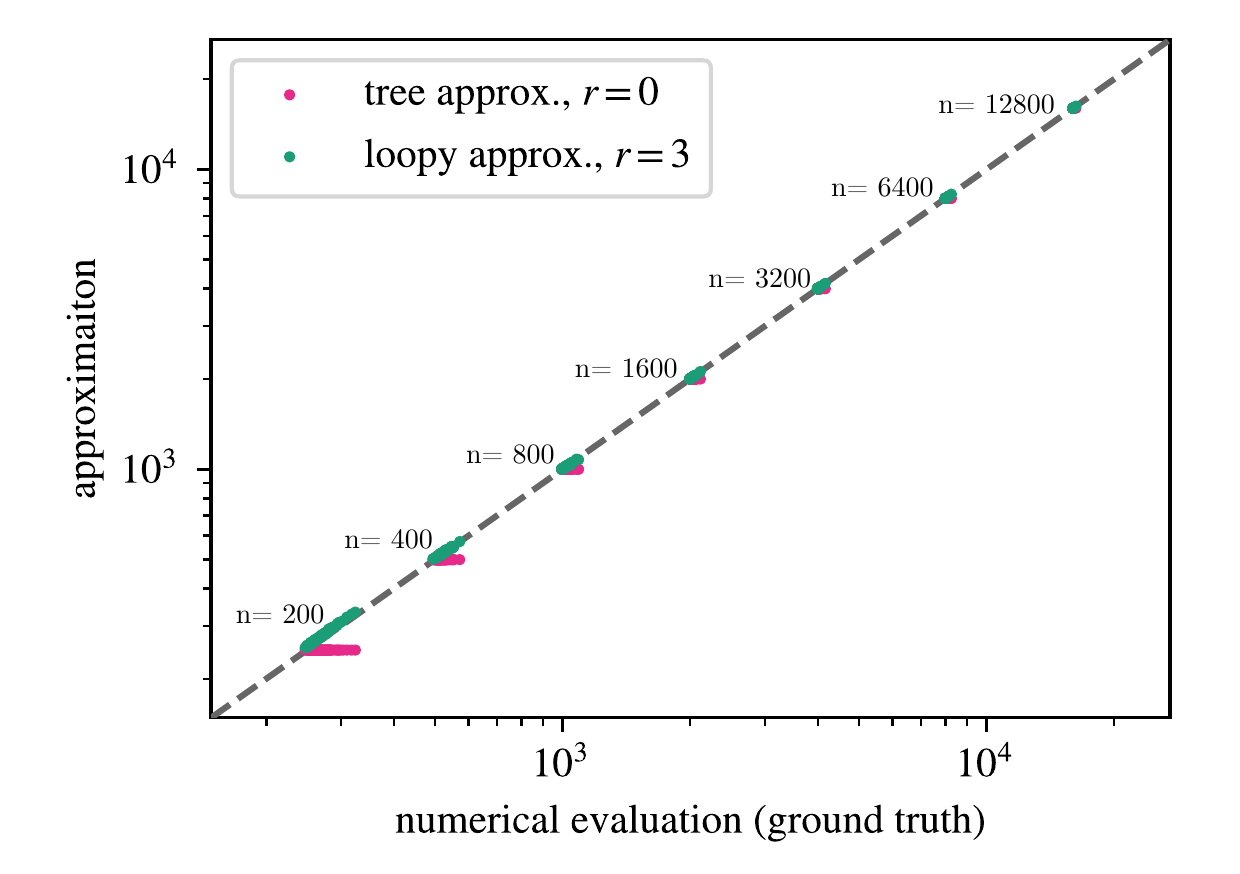}
	\caption{First return time distribution approximations against ground truth, for regular stochastic block models of varying size but with a fixed community structure.
 The exponential tail of randomly chosen nodes in networks of differing sizes was compared to approximations.
 As the number of nodes increases, the tree approximation moves towards the line $y=x$, showing that the approximation is converging to the truth.
 For the $r=3$ approximation, the points are close to the line $y=x$ even for small $n$.
 }
  \label{fig:cc}
\end{figure}

In Fig.~\ref{fig:cc}, we compare the tail slope predicted by Eq.~\eqref{eq:tail_mean} to those found by numerically iterating the random walk matrix.
These networks were generated by the degree regular stochastic block model sampler of graph-tool \cite{peixoto_graph-tool_2014}.
Each network has $20$ communities and each node has degree $6$.
The community structure is relatively strong. 
Parameters were set so that, on average, $2/3$rds of the neighbours of any node will be in the same group as the node itself.
But, as $n$ increases we see the approximation converges to the ground truth values.
Still, one should not conclude that the tree approximation is good for all networks.
For instance, in Fig.~\ref{fig:cc} we see that the approximation is less accurate for small networks.
This, we believe, is due to the presence of many short cycles in the graphs.

To further investigate how the approximation fails, we look at the case in which the number of communities grows with the size of the network, specifically $c = n/20$.
In this case, the networks are not locally tree-like, and so it is not surprising that the properties of the random walks  are significantly different from those on random graphs.
Figure~\ref{fig:sc} shows the results of these experiments.
In this case we see that even as $n$ increases, the approximation fails to converge to the ground truth.

In the next section, we correct the approximation to account for short cycles and thus correct for the issues we observe with the tree approximation.
With this, we derive an approximation that we believe should be accurate in a very wide number of cases.

\subsection{Approximation for networks with short cycles}\label{sec:ca_approx}

In order to derive an approximation for networks with short cycles, we follow refs.~\cite{cantwell2019message, cantwell2023heterogeneous} and first define the $r$-neighbourhood around each node.
For a given integer $r$, the $r$-neighbourhood of node~$i$ consists of all of the nodes immediately adjacent to node~$i$, along with all edges and all nodes on paths of length $r$ or shorter between nodes adjacent to $i$.
We denote the $r$-neighbourhood of node $i$ as $\mathcal{N}_i^{(r)}$.

Now, we are in a position to write the message-passing approximation, following a similar argument to before.
Any excursion from node~$i$ will first move to an immediate neighbour. 
Call this neighbour $j$.
Once at $j$, the excursion will leave the neighbourhood and return to $j$ some $m \geq 0 $ times.
After this, it will either immediately return to $i$, or move on to some other node in the neighbourhood.

In effect, each excursion from node~$i$ can be broken down into two parts: (i) an excursion in the neighbourhood of $i$, and (ii) a number of excursions from each node in the neighbourhood.
An excursion from $i$ in the neighbourhood $N_i^{(r)}$ can be denoted $w = i \to j \to \dots \to i$.
Let the set of all excursions in the neighbourhood of $i$ be labeled $W_i$.
Then the probability of returning at time $t$ is
\begin{equation}
	\widetilde{Y}_i(t) = \frac{1}{k_i} \sum_{w \in W_i} \sum_{t_j : j \in w} \delta(t, l+1+{\textstyle \sum_{j \in w}} t_j) \prod_{j \in w} \frac{ X_{i \leftarrow j}(t_j) }{k_j}
\end{equation}
where $l$ is the length of the excursion $w$, and $P(w)$ is the probability of the excursion (i.e. the product of the transition probabilities).

From this, we now compute the generating function
\begin{align}
	\widetilde{F}_i(z) &= \sum_{t=1}^{\infty} \widetilde{Y}_i(t) z^t \nonumber \\ 
	&= \frac{1}{k_i} \sum_{w \in W_i}  \sum_{ \{ t_j : j \in w \} }  z^{l+1} \prod_{j \in w} X_{i \leftarrow j}(t_j) z^{t_j} / k_j \nonumber \\
	&= \frac{z}{k_i} \sum_{w \in W_i} \prod_{j \in w} \frac{z}{k_j - k_j \widetilde{F}_{i \leftarrow j}(z) }.
	\label{eq:r_marginal}
\end{align}
By the same arguments,
\begin{equation}
	\widetilde{F}_{i \leftarrow j}(z) = \frac{z}{k_j} \sum_{w \in W_{j \setminus i}}  \prod_{k \in w} \frac{z}{k_k  - k_k \widetilde{F}_{j \leftarrow k}(z)}.
	\label{eq:r_message}
\end{equation}
The sum over all walks can be computed by matrix inversion.
Define
\begin{equation}
	u_{j}^{i} = \frac{A_{ij}}{k_i },
\end{equation}
\begin{equation}
	v_{j}^{i}(z) = \frac{ A_{j i}}{k_j-k_j \widetilde{F}_{i \to j}(z) },
\end{equation}
and
\begin{equation}
	B_{jk}^{i}(z) = \frac{ A_{j k} }{ k_j-k_j \widetilde{F}_{i \to j}(z) },
\end{equation}
then Eq.~\eqref{eq:r_marginal} becomes
\begin{equation}
	\widetilde{F}_i(z) = \boldsymbol{u}^i \big( 1  - B^{i}(z) \big)^{-1} \boldsymbol{v}^i(z) .
\end{equation}

However, again, Eqs.~\eqref{eq:r_marginal} and~\eqref{eq:r_message} do not meet the requirements that $\widetilde{F}_i(1)=1$ or $\widetilde{F}_i'(1)=2m/k_i$.
We propose making the same adjustment as before: adding the generating function of a geometric distribution with an appropriately defined mean.
In other words, letting 
\begin{equation}
	h_i = \frac{2 m  - k_i \widetilde{F}_i'(1) }{ k_i - k_i \widetilde{F}_i(1) }
\end{equation}
we approximate $F_i(z)$ as
\begin{equation}\label{eq:final_approx}
	F_i(z) = \widetilde{F}_i(z) + \frac{ z - z \widetilde{F}_i(1)}{z+h_i -z h_i}
\end{equation}

\subsection{Numerical experiments for graphs with short cycles}\label{sec:cycles_sims}

\begin{figure}\
  \includegraphics[width=0.45\textwidth]{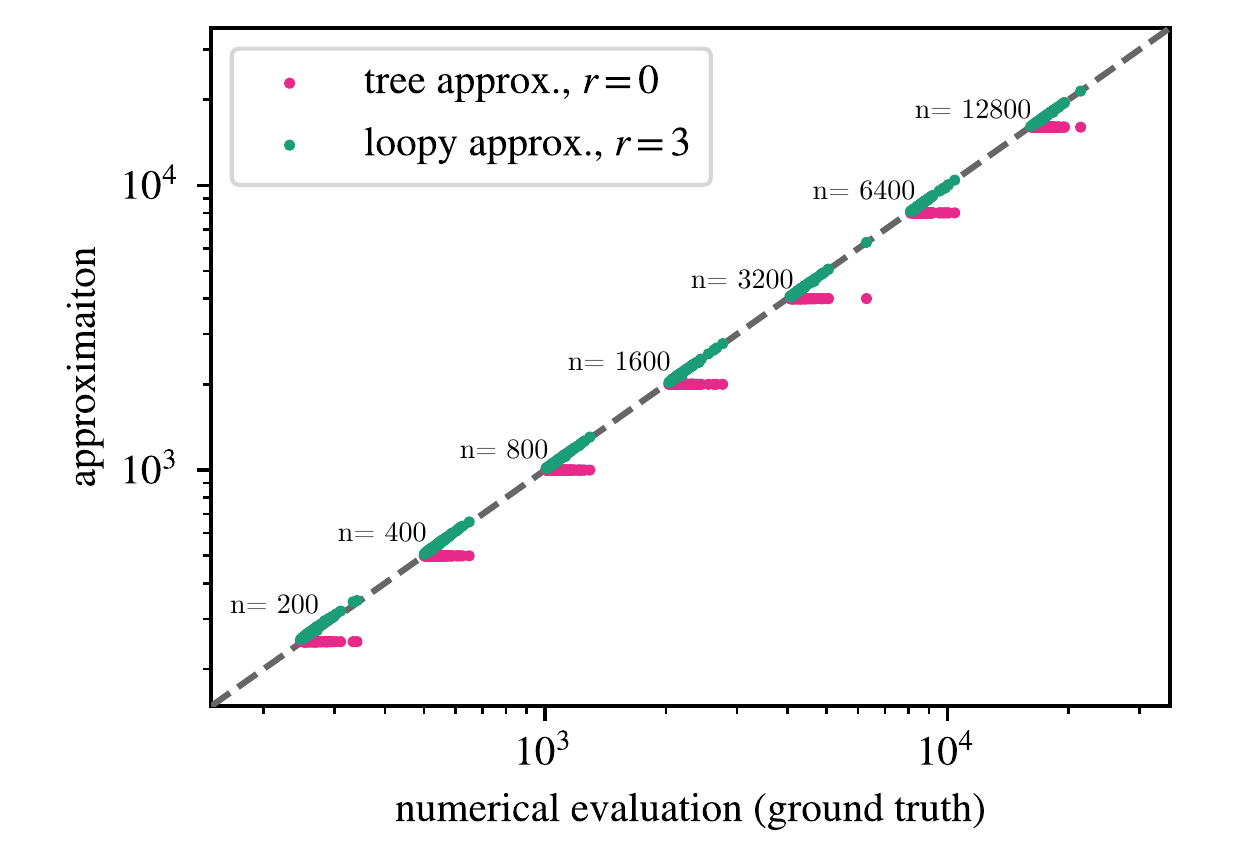}
	\caption{First return time distribution approximation against truth for regular stochastic block models of varhous sizes variable number of communities $c = n/20$.
The exponential tail of randomly chosen nodes in networks of differing sizes was compared to approximations.
The predictions for the tree approximation are not very close to the line $y=x$, even as $n$ increases.
Conversely, the predictions for $r=3$ remain close to $y=x$ for all $n$.
 }
  \label{fig:sc}
\end{figure}

Returning to Fig.~\ref{fig:cc}, we see that the combined approximation looks accurate even for smaller $n$.
For large $n$, the ground truth and both approximations of Eqs.~\eqref{eq:finite_approx} and \eqref{eq:final_approx} converge to a point on the line $y=x$.
However, for smaller $n$, where there is a significant variance due to finite size effects, only the improved approximation of Eq.~\eqref{eq:final_approx} correctly falls on the line $y=x$.
In Fig.~\ref{fig:sc}, where the number of communities grows linearly with network size, we observe the discrepancy between the approximation of Eq.~\eqref{eq:finite_approx} for all $n$.
In contrast, we see that the approximation of Eq.~\eqref{eq:final_approx} is good for all $n$.

\section{Discussion}\label{sec:discussion}

Random walks are influenced by the structure of the underlying network. It is remarkable that the mean return time has such a simple formula: $\mu(i \leadsto i) = 2m/k_i$.
This formula depends on only two quantities.
First, the degree of node~$i$, $k_i$, which is an entirely local property.
And second, twice the number of edges, $2m$, a basic global property.
Our approximations are inspired by a similar line of thinking.
We relate the full distribution of returns to a mixture of two processes: a random walk in the local neighbourhood, and a global mean-field process.
To do so, we have developed a message passing formalism to capture the local walks, and have corrected it with the predictions of the global mean-field process, using Eq.~\eqref{eq:tail_mean}. The only global property used in the correction is the total number of edges, $m$.
An integer-valued parameter $r \geq 0$ controls the accuracy of our approximations on networks that are not locally tree-like, and we find good agreement for small values of $r$ in our experiments.
Overall, we see good results both at short times and asymptotically.

While the approach appears to work well, we note that it is derived from intuitive arguments only.
We do not see a fully rigorous justification or proof for why the approximation must work in practice.
As a result, the evidence for the effectiveness of the approach is numerical rather than analytic.
Hence, these findings elicit further questions.
Why does this approach work so well?
The story is a similar one to other intuitive, physics-derived, approximations.
For example, it is known that the mean-field approximation to the Ising model is premised on an assumption that is known to be false---that each spin feels only the average effect of the system \cite{gleeson2012accuracy}.
Nevertheless, this \emph{wrong} assumption provides an accurate approximation, particularly in high dimensions.
While the approximation can be justified through its accuracy, it has also prompted the community to ask \emph{why} it works \cite{albert2002statistical}.
Similarly, researchers have investigated the reasons why tree-based methods for dynamical processes on networks work ``unreasonably" well even in
networks with clustering \cite{melnik2011unreasonable}.

At a high level, the fact that our random walk approximations is accurate suggests that large-scale structures do not greatly influence the return times of random walks, at least for random graphs models.
Interestingly, no matter how strong the communities are in a stochastic block model, our result says the return statistics are the same as an Erdős–Rényi graph, so long as the number of groups is fixed as $n \to \infty$.
This is indeed what we found in Fig.~\ref{fig:cc}.
A recent work of L{\"o}we and Terveer \cite{lowe2024hitting}, proves a concordant result, that the average hitting time in stochastic block models looks the same as for Erdős–Rényi graphs under appropriate conditions.
This observation seems concerning for community detection methods that rely on the statistics of finite-length random walks.
In contrast, the networks in Fig.~\ref{fig:sc} have a number of communities that scales with the size on the network. Thus, the limit is not locally tree-like, and the random walk statistics are always easily distinguished from the Erdős–Rényi graph.

In summary, we derived an approximation for the statistics of random walks.
We saw that the statistics are strongly affected by cycles and local structure, and incorporated that fact into our approximations.
On the other hand, the global structure only appeared to influence the statistics in a fairly superficial way---only through the quantity $2m$, i.e. twice the number of edges in the network.
We looked at walks on undirected and unweighted networks.
However, we see no reason that the approach does not generalize to the weighted and directed case, i.e., to arbitrary Markov chains on finite spaces.
This, however, we leave to future work.

 \begin{acknowledgments}
R.L. acknowledges support from the EPSRC Grants EP/V013068/1 and EP/V03474X/1, and G.T.C from NSF Grant BIGDATA-1838251.
 \end{acknowledgments}


\appendix
\section{Population dynamics algorithm}\label{sec:pop_dyn_alg}

Here, we describe how to numerically solve Eq.~\eqref{eq:pop_dyn_message} to estimate $P_{\leftarrow}^{(z)}(F, k)$.
We do this using a Markov chain Monte Carlo method, known as population dynamics \cite{mezard2009information}.

First, for the sake of argument, suppose we already had a set of $N$ samples, $Q = \big(F_i, k_i \big)_{i=1\dots N}$.
For now, do not worry about how we obtained these samples.
Assuming $Q$ is correctly sampled from $P_{\leftarrow}^{(z)}$, then we could generate a new sample as follows.
First, generate a degree $k$ proportional to $k \rho_k$, then take $k$ randomly chosen members of $Q$, call these $(F_j, k_j)$, and set $F = \sum_{j=1}^{k} \frac{z^2}{k k_j (1-F_j)}$.
After doing this, the pair $(F, k)$ will also be a correct sample from $P_{\leftarrow}^{(z)}$.
This is the key observation behind the population dynamic algorithm.

In the population dynamic algorithm, we begin with an initially incorrect set of samples---we initialize $Q$ to be sampled from some other tractable distribution, e.g., a uniform distribution.
Then, we create new (also initially incorrect) samples by using the above process, and replace the samples in $Q$ with the new samples.
A full sweep of this process is as follows:
\begin{algorithm}
for $i=1, \dots, N $:
    $F \gets 0$
    $k \sim k \rho_k / \langle k \rangle$
    for $u = 1, \dots, k$:
        $j \sim \text{Uniform}(N)$
        $F \gets F + \frac{z^2}{k k_j (1- F_j)}$
    $(F_i, k_i) \gets (F, k)$
\end{algorithm}
As stated previously, if $Q$ were a correct set of samples, the above sweep would leave us with a new set of correct samples.
Or, in other words, a correct set of samples is the stationary distribution of this sweep process.
Further, only distributions that solve Eq.~\eqref{eq:pop_dyn_message} will be stationary under this process.
So, starting out from an arbitrary distribution, we simply need to run a large number of sweeps to reach the stationary distribution, and thus numerically sample the distribution of interest,~$P_{\leftarrow}^{(z)}$.

\end{document}